%% file: aaai.tex
\documentclass[letterpaper]{article}
\usepackage{aaai}
\usepackage{times}
\usepackage{helvet}
\usepackage{courier}
\usepackage{graphicx}
\usepackage{caption}
\usepackage{subcaption}
\usepackage{booktabs}
\usepackage{blindtext}
\usepackage{amsmath}
\usepackage{multirow}
\usepackage{color}
\usepackage{url}

\begin{document}
%
\title{Causal Feature Selection for Individual Characteristics Prediction}

\author{Tao Ding$^*$, Cheng Zhang$^\dagger$, \and Maarten Bos$^\dagger$ \\
       $*$ Department of Information Systems \\
       University of Maryland, Baltimore County \\ 
       \texttt{taoding01@umbc.edu}
       \thanks{This work was accomplished during the period of the first author working as research intern in Disney Research.}\\
       $\dagger$Disney Research,\\
       Pittsburgh, USA\\
       \texttt{\{cheng.zhang,mbos\}@disneyresearch.com}
       }
       
\maketitle

\begin{abstract}
\begin{quote}
People can be characterized by their demographic information and personality traits. Characterizing people accurately can help predict their preferences, and aid recommendations and advertising. A growing number of studies infer people’s characteristics from behavioral data. However, context factors make behavioral data noisy, making these data harder to use for predictive analytics. In this paper, we demonstrate how to employ causal identification on feature selection and how to predict individuals' characteristics based on these selected features. We use visitors' choice data from a large theme park, combined with personality measurements, to investigate the causal relationship between visitors’ characteristics and their choices in the park. We demonstrate the benefit of feature selection based on causal identification in a supervised prediction task for individual characteristics. Based on our evaluation, our models that trained with features selected based on causal identification outperformed existing methods. 
\end{quote}
\end{abstract}
\input{intro}
\input{related}

\input{data}
\input{causal}

\input{exp_pred}
\input{result_analysis}
\input{discussion}
\input{acknowledgement}
\bibliographystyle{named}

{\footnotesize
\bibliography{aaai}
}

\end{document}

%% file: intro.tex
\section{Introduction}
Understanding an individual's characteristics is useful for many real life applications. The term individual characteristics refers to individual differences in characteristic patterns of thinking \cite{widiger2011personality}, and includes both demographics and personality. Knowing someone's individual characteristics can help to understand that person's preferences, which has important applications ranging from health care (Smith and Spiro III 2002; Giota and Kleftaras 2013) 
to marketing \cite{spence1997use,odekerken2003strengthening,jamal2001consumers,lin2002segmenting,wang2013text,yang2015using,ding2016personalized}, health \cite{booth1994associations,smith2002personality,giota2013role} and public politics \cite{greenstein1992can,caprara2006personality,chirumbolo2010personality}. Individual characteristics -- such as gender and personality - are relatively stable over time. E.g. the people who display higher score of a particular personality trait 
continue to display higher score of this personality trait over time, when compared to other individuals \cite{caspi2005personality}. This means that it is practical, and relatively stable over time, to predict a person's behavior and preferences from individual characteristics. 

Personality information can be arduous to obtain.
A traditional approach to measuring personality requires participants to take a psychological test (e.g. filling out questionnaires), which is time-consuming and difficult to scale up. It is therefore desirable to circumvent this testing process and instead directly predict individual characteristics based on readily available observational data.

\begin{figure}[t]
\includegraphics[width=0.45\textwidth]{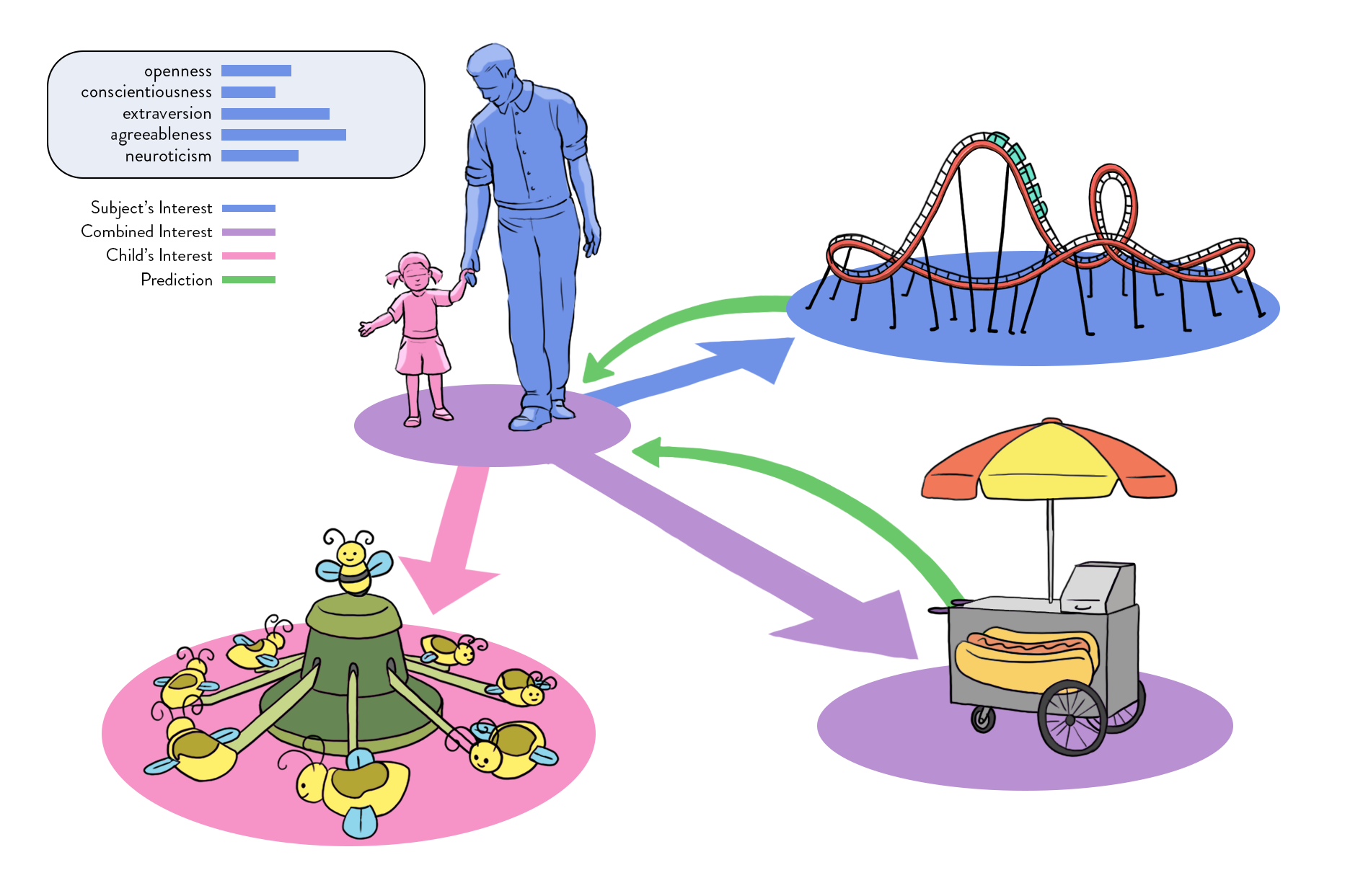}
\caption{An exemple of individual characteristics prediction using theme park activities. The activities are caused by both the target individual's preferences and the contextual situation, which in this case is the accompanying child. To separate the signal from noise for the individual characteristics prediction task, we propose to use a causal identification method to determine the informative activity for this task.}
\label{fig:intro}
\end{figure}

A growing proportion of human activities, such as consuming entertainment, shopping, and social interactions, are now mediated by digital services and devices. These digital footprints of consumer behaviors can be recorded and analyzed. Understandably, there is an interest in automatically predicting individual characteristics including age, gender, income and personality traits from these digital footprints. In this work, we focus on predicting individual characteristics from real-life experiences in theme parks. 

This task is challenging because behavioral data are noisy. Noisiness can be caused by different factors, such as that while people move in groups, we are trying to predict personality at an individual level \cite{aronson2012social,zhang2015groupbox}. In our study we use behavioral data, such as which facilities a person visited during their stay in the theme park, to predict traits such as openness or agreeableness. Imagine a person is visiting a theme park with his/her child, as shown in Figure \ref{fig:intro}, and our goal is to predict the parent's traits. The activities of the parent are influenced by the fact that the child is with them. For example, a visit to a toy shop may not be the result of the parent's traits, but of the context factor (the child). The signal of the parent's personality is partially drowned out by the noise of the context. This makes it challenging to distinguish individual differences from group behaviors. Identifying which behaviors are directly caused by a person's characteristics is essential.





In the paper, we propose a new machine learning pipeline to predict individual characteristics. We present a study on 3294 visitor profiles of a large theme park. We first employ causal identification to identify individual behavioral features. Second, we use these features to build predictive models to predict specific personality traits. Our contribution is threefold:
\begin{enumerate}
\item This is the first study that investigates the relationship between someone's location history (defined as visits of facilities in a theme park) and their personality.
\item We take advantage of causal identification methods for feature selection and build a prediction model to predict individual characteristics. Our results demonstrate improvement over baseline models, such as models that use all features or models that use traditional feature selection methods.  
\item Compared to other machine learning methods, causal identification is able to obtain interpretable results in a principled way and provide more insights in behavior. 
\end{enumerate}

%% file: related.tex
\section{Background and Related Works}

We aim to predict individual characteristics from real-life theme park visits data with causal identification for feature selection. Therefore we first explain why individual characteristics are interesting and important. Then we summarize related work on individual characteristics prediction using various type of data. Finally we discuss research on causal identification and causal inference.

\paragraph{Individual characteristics} 
We define individual characteristics as both demographic information and personality traits. In the context of  our application, demographic information includes age, income, and the number of kids included in a theme park visit; personality traits are described using the well established Big5 personality model \cite{goldberg1993structure}.
The Big5 personality model \cite{goldberg1993structure} represents personality with scores on five personality traits. These traits are \textit{openness}, \textit{conscientiousness}, \textit{extraversion}, \textit{agreeableness} and \textit{neuroticism}. Table \ref{tab:big5personality} shows a description of the traits. The Big5 model is widely used to represent a person's personality.


There is a rich body of work in behavioral science on the relationship between humans' individual characteristics and their real-world behaviors. 

One example study showed that age, level of education, and income among 250 hotel restaurant customers was correlated with complaint behavior \cite{sujithamrak2005relationship}. Another study revealed a relationship between smoking behavior and demographic variables \cite{moody1980relationships}.
\begin{figure*}
\caption{\bf The Distribution of Big 5 Personality traits.}
\centering
\captionsetup{justification=centering}
\begin{subfigure}[b]{0.19\textwidth}
\centering
\captionsetup{justification=centering}
\includegraphics[width=.95\textwidth]{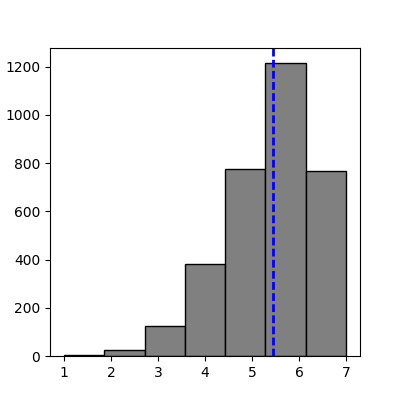}
\caption{Agreeableness}
\end{subfigure}
\begin{subfigure}[b]{0.19\textwidth}
\centering
\captionsetup{justification=centering}
\includegraphics[width=.95\textwidth]{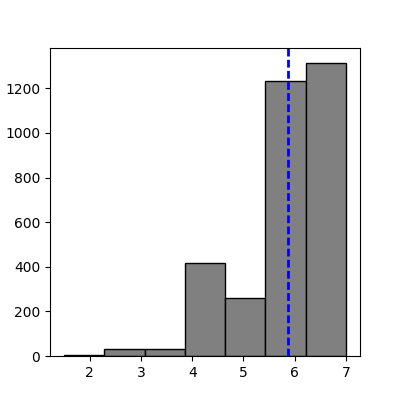}
\caption{Conscientiousness}
\end{subfigure}
\begin{subfigure}[b]{0.19\textwidth}
\centering
\captionsetup{justification=centering}
\includegraphics[width=.95\textwidth]{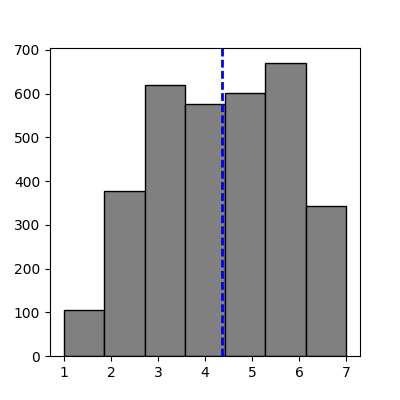}
\caption{Extraversion}
\end{subfigure}
\begin{subfigure}[b]{0.19\textwidth}
\centering
\captionsetup{justification=centering}
\includegraphics[width=.95\textwidth]{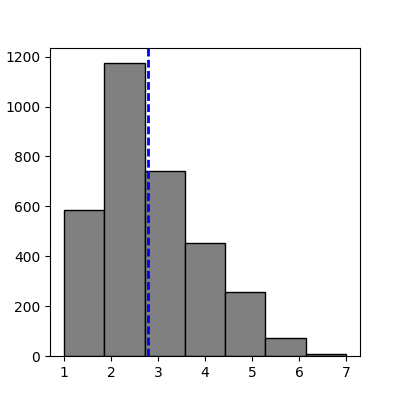}
\caption{Neuroticism}
\end{subfigure}
\begin{subfigure}[b]{0.19\textwidth}
\centering
\captionsetup{justification=centering}
\includegraphics[width=.95\textwidth]{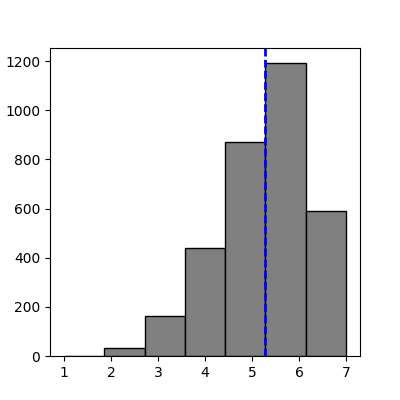}
\caption{Openness}
\end{subfigure}
\label{fig_big5}
\end{figure*}
Besides demographics, personality traits also predict behavior. One study found that smokers score higher on openness to experience and lower on conscientiousness, a personality trait related to a tendency to show self-discipline, act dutifully, and aim for achievement \cite{campbell2014personality}. A second study examined the link between personality and alcohol consumption and found that alcohol use is correlated positively with sociability and extraversion \cite{cook1998personality}.
Another study showed a link between personal environments and personality \cite{gosling2002room}. Yet another study \cite{fennis2007you} found that a brand's personality profile may carry over and affect perceptions of the personality of the brand's owner. These studies are often carried out using questionnaires, surveys, or interviews. These methods are time-consuming, and a more efficient way to predict traits would be useful.

\begin{table}[t]
    \caption{Big Five personality model \cite{ding2016personalized}}
    \label{tab:big5personality}
    \begin{tabular}{ l l} 
    \toprule
    \textbf{Personality}& \textbf{Description}\\
    \cmidrule(lr){1-2}
Openness& \begin{minipage}{5cm}A person's level of intellectual curiosity, creativity and preference for novelty and variety. \end{minipage}\\
\cmidrule(lr){1-2}
Conscientiousness & \begin{minipage}{5cm} A person's tendency to be organized and dependable, show self-discipline, act dutifully, and prefer planned rather than spontaneous behavior.\end{minipage}\\
\cmidrule(lr){1-2}
Extraversion & \begin{minipage}{5cm} A person's energy, positivity, assertiveness, sociability, talkativeness, and tendency to seek stimulation from the company of others.\end{minipage}\\
\cmidrule(lr){1-2}
Agreeableness & \begin{minipage}{5cm}A person's tendency to be compassionate and cooperative. Also a measure of one's trustingness, helpfulness, and well-tempered nature.\end{minipage}\\
\cmidrule(lr){1-2}
Neuroticism &\begin{minipage}{5cm} A person's tendency to experience unpleasant emotions easily, and have low emotional stability and impulse control.\end{minipage}\\
 \bottomrule
    \end{tabular}
  \end{table}
  
 \paragraph{Machine Learning for Individual Characteristics Prediction}
Using machine learning methods to predict individual characteristics from behaviors has increasingly gained attention. Ideally, various types of real-life behavior data would be used to predict individual characteristics. However, existing research is mainly limited to the use of social network data to predict individual characteristics.

There have been several studies to show that users' digital footprints on social networks can be used to infer their demographics and personalities \cite{farnadi2013recognising,mislove2011understanding,golbeck2011predicting,kosinski2013private}. When there is a large volume of data available – especially for high dimensional data – feature learning is needed to reduce data dimensionality \cite{kosinski2015facebook,pennebaker2007linguistic,kosinski2013private,schwartz2013personality}. 

An alternative to this approach is supervised learning. Studies often focus on recognizing individual characteristics from a limited amount of noisy annotated data, using univariate and multivariate regression formulations. Supervised learning can be used either on the raw data or on extracted features. These features can be extracted using representation learning methods.
In supervised learning, feature selection can improve the accuracy of the learning algorithm. There are existing methods to identify relevant features and remove irrelevant and redundant features before training a model. In a regression model, the common method is to apply L1 regularization that adds a penalty $\alpha \sum_{n}^{i=1} |w_{i}| $ to the loss function (L1-norm), which forces weak features to have zeros as coefficients. This is also called LASSO regression \cite{tibshirani1996regression}. This inherently creates feature selection. Another method is correlation based feature selection (CFS) which filters the features by performing feature selection based on correlation analysis \cite{yu2003feature}. However, both these methods do not provide interpretable results and lack principled justification.

\paragraph{Causal Identification} A fundamental assumption for research on individual characteristics prediction is the causal relationship between the individual characteristics and behaviors.  Additionally, prediction is only possible if the target is the cause of the data (effect) \cite{scholkopf2012causal}.
Most feature selection methods do not attempt to uncover cause-effect relationships between feature and target \cite{guyon2007causal}. A correlated feature may be predictive of an outcome without intervention. Causal identification \cite{spirtes2000causation} can identify features with direct effects caused by target variables which contain predictive information of the target. Additionally, the result from causal identification - a graph indicating causal relationships - is highly interpretable. In individual characteristics identification, causal identification is able to determine which behaviors are caused by individual characteristics and which behaviors are caused by the context. 
In this paper, we apply different causal identification methods to identify direct effects caused by personal traits. Using causal identification methods for feature selection leads to better predictive performance in our application.

%% file: data.tex
\section{Dataset}
The data for our study were collected from visitors of a large theme park resort. The resort has over 30 hotels, more than 100 restaurants and hundreds of attractions. We assume individual characteristics affect the choices visitors make. We asked visitors' permission to use their choice and location history. From visitors who agreed to this we retrieved the locations they visited, including restaurants, stands and kiosks, attractions\&rides, stores and other entertainment. Participating visitors also allowed us access to data such as their park entry time, hotel check-in time, purchases, and other metadata such as length of stay, trip cost, and which of the parks they visited first (see Table \ref{tab:metadata}).


\begin{table}[t]
    \caption{The Meta-data of User's Profile}
    \label{tab:metadata}
    \begin{tabular}{ p{2.2cm} p{2cm} p{2.3cm}} 
     \toprule
\textbf{Item}	&	\textbf{Data Type}  & \textbf{Representation}\\
\cmidrule(lr){1-3}
Gender	&	Categorical & 0: male 1: female\\
\hline
Age	&	Categorical & 17-78\\
\hline
Total visits & Numerical &0-99\\
\hline
Vacation days & Numerical &0-99 \\
\hline
Income & Numerical & 1-14\\
\hline
Trip cost & Numerical & 0-9999\\
\hline
\# of children & Numerical & 0-9\\
\hline
Hotel choice & Categorical & 0: No 1: Yes\\
\hline
How the trip was booked & Categorical & 13 categories \\
\hline
Dining plan & Categorical & 0: No 1: Yes\\
\hline
First park visited & Categorical &  4 different theme parks\\
\hline
\end{tabular}
\end{table}

\begin{figure}[h]
\caption{\bf The Distribution of Demographics}
\centering
\hspace{-25pt}
\begin{subfigure}[b]{0.17\textwidth}
\centering
\captionsetup{justification=centering}
\includegraphics[width=.9\textwidth]{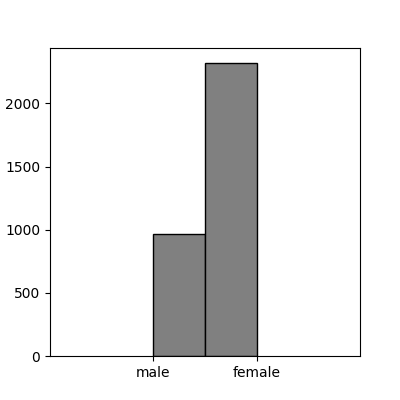}
\caption{Gender}
\end{subfigure}
\hspace{-15pt}
\begin{subfigure}[b]{0.17\textwidth}
\centering
\captionsetup{justification=centering}
\includegraphics[width=.9\textwidth]{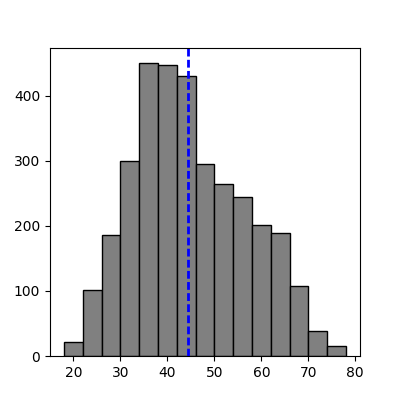}
\caption{Age}
\end{subfigure}
\hspace{-15pt}
\begin{subfigure}[b]{0.17\textwidth}
\centering
\includegraphics[width=.9\textwidth]{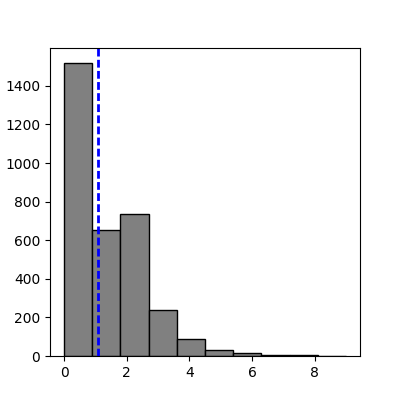}
\captionsetup{justification=centering}
\caption{\# of children}
\end{subfigure}
\hspace{-25pt}
\label{fig_demog}
\end{figure}

\begin{figure*}[htb]
\caption{\bf Stability of Feature Selection for Big5 Personality.}
\centering
\captionsetup{justification=centering}
\hspace{-20pt}
\begin{subfigure}[b]{0.21\textwidth}
\centering
\captionsetup{justification=centering}
\includegraphics[width=.85\textwidth]{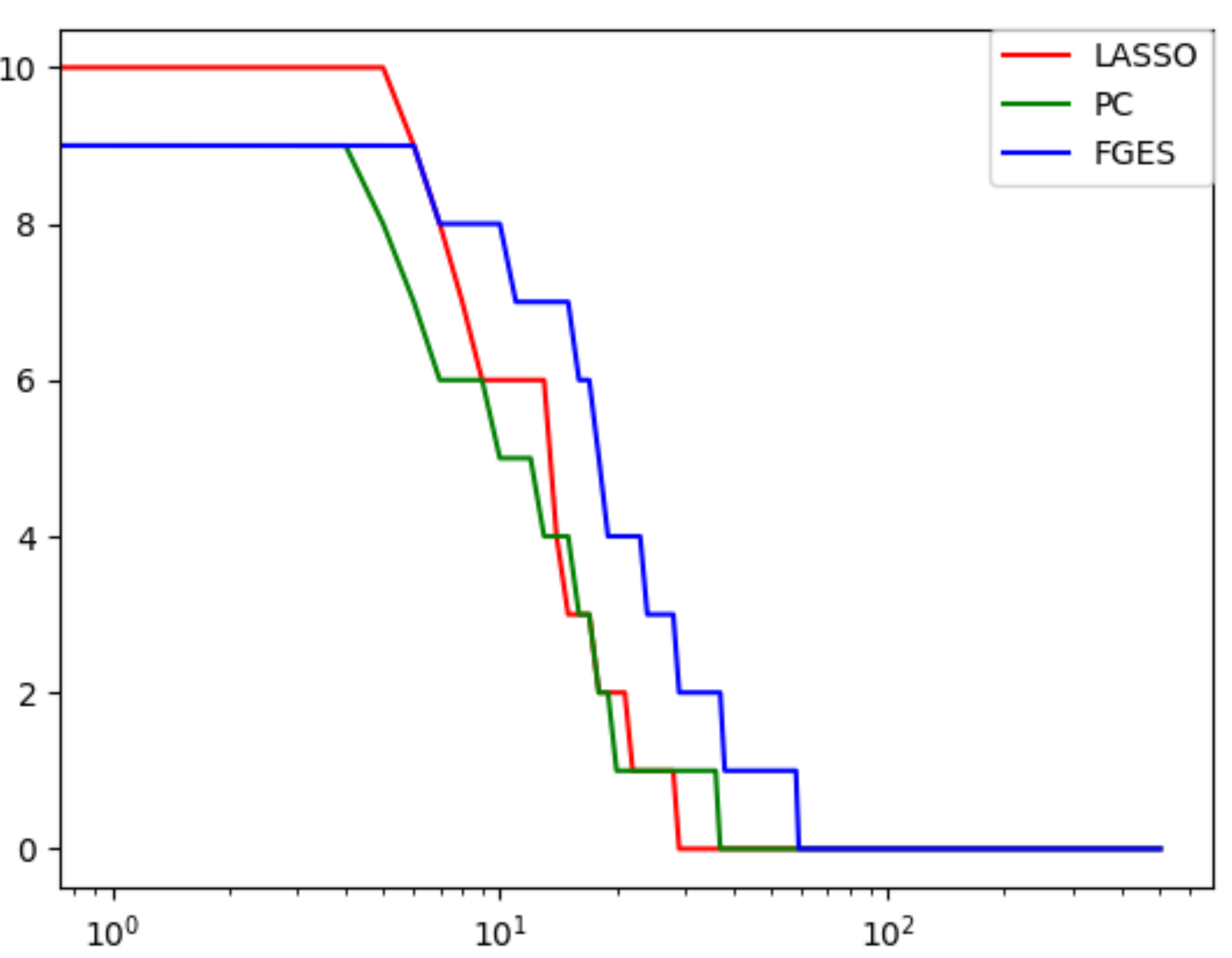}
\caption{Agreeableness}
\end{subfigure}
\hspace{-15pt}
\begin{subfigure}[b]{0.21\textwidth}
\centering
\captionsetup{justification=centering}
\includegraphics[width=.85\textwidth]{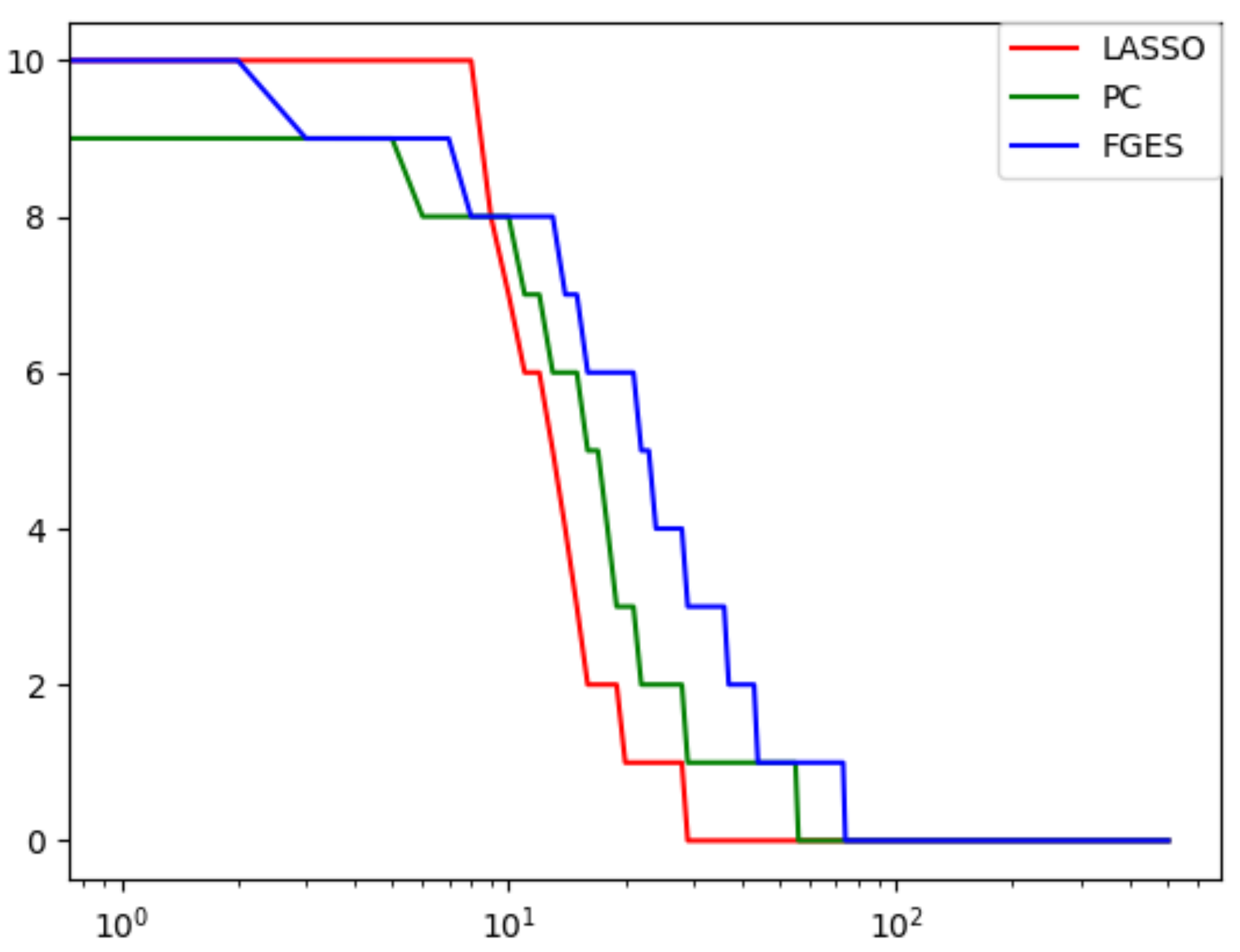}
\caption{Conscientiousness}
\end{subfigure}
\hspace{-15pt}
\begin{subfigure}[b]{0.21\textwidth}
\centering
\captionsetup{justification=centering}
\includegraphics[width=.85\textwidth]{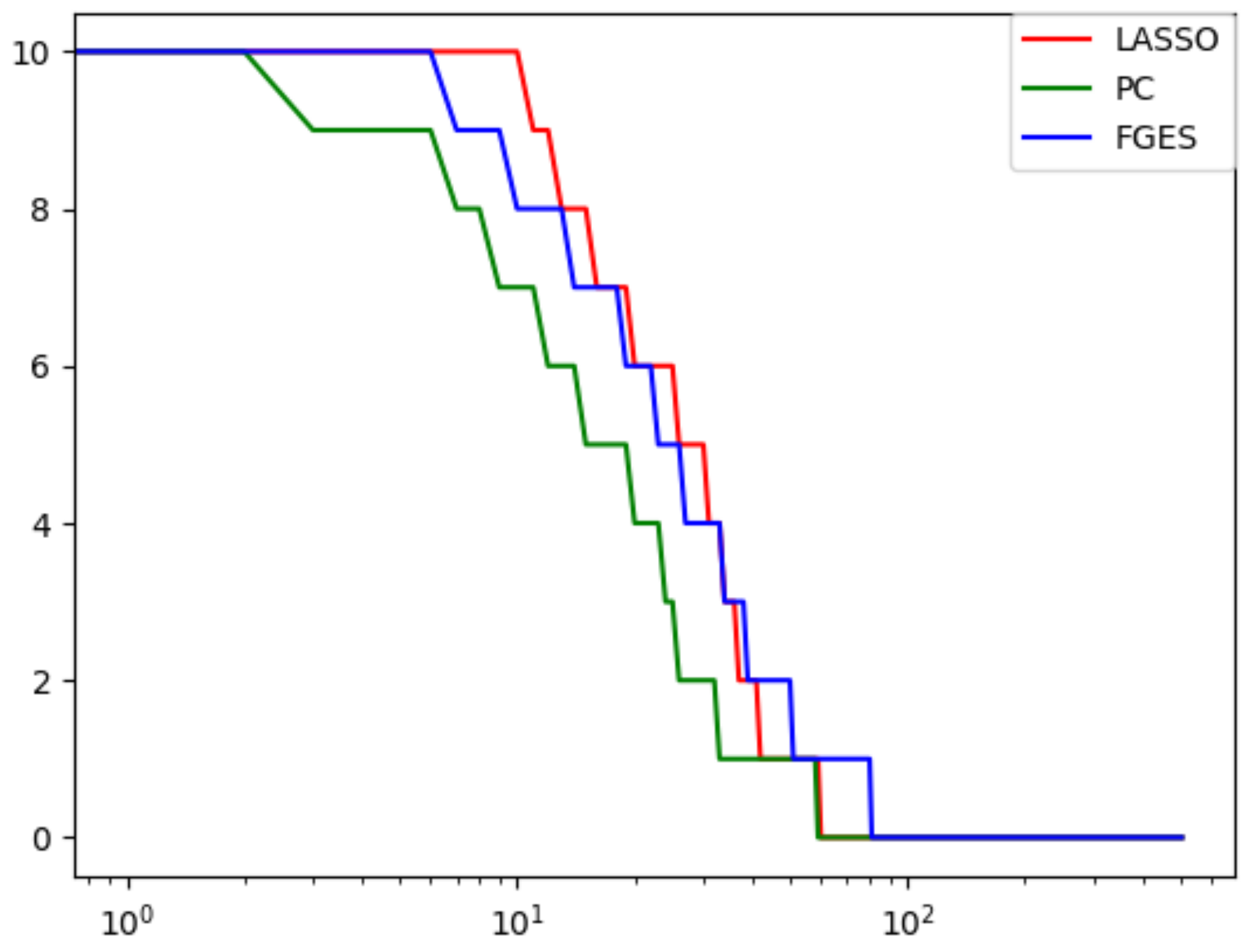}
\caption{Extraversion}
\end{subfigure}
\hspace{-15pt}
\begin{subfigure}[b]{0.21\textwidth}
\centering
\captionsetup{justification=centering}
\includegraphics[width=.85\textwidth]{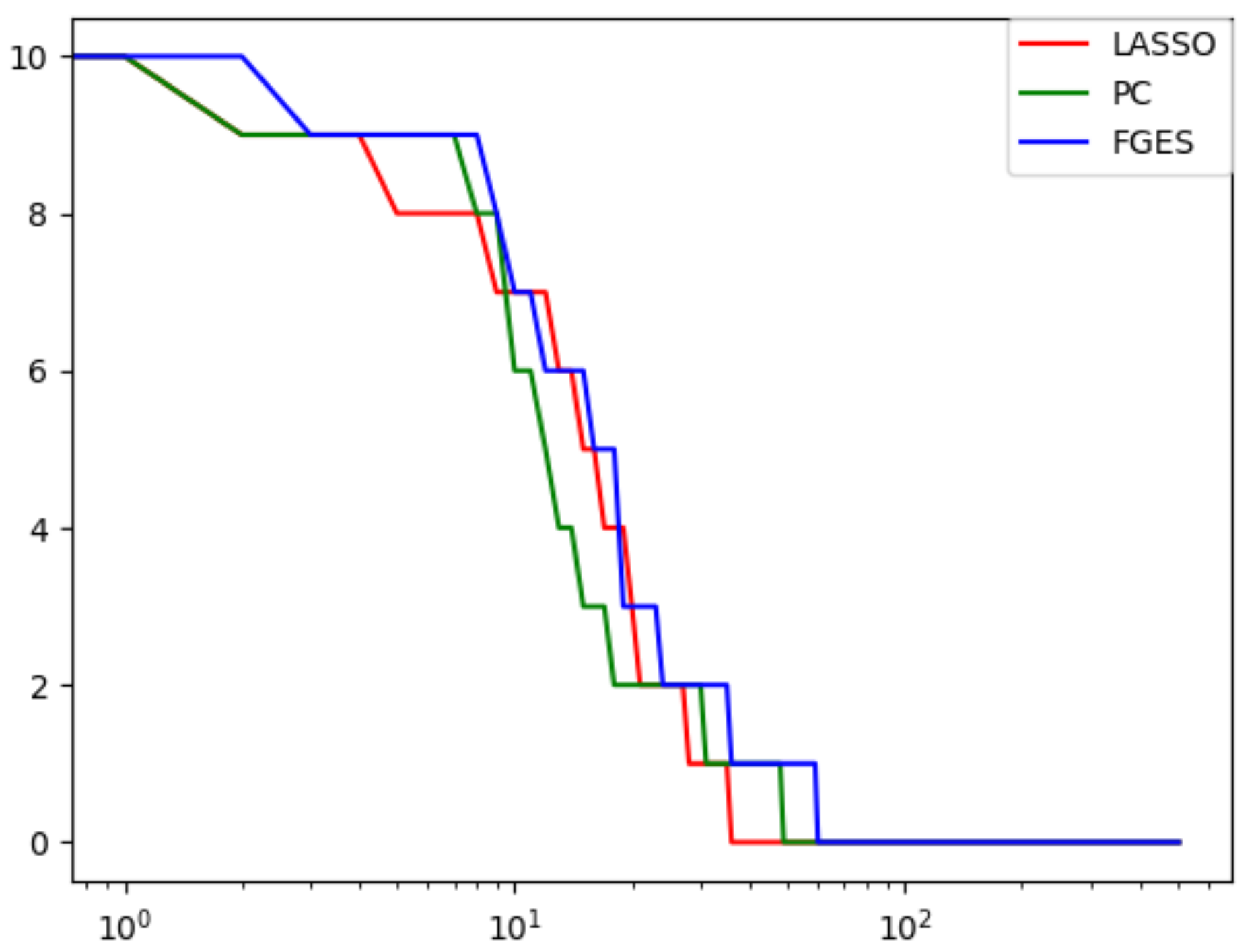}
\caption{Neuroticism}
\end{subfigure}
\hspace{-15pt}
\begin{subfigure}[b]{0.21\textwidth}
\centering
\captionsetup{justification=centering}
\includegraphics[width=.85\textwidth]{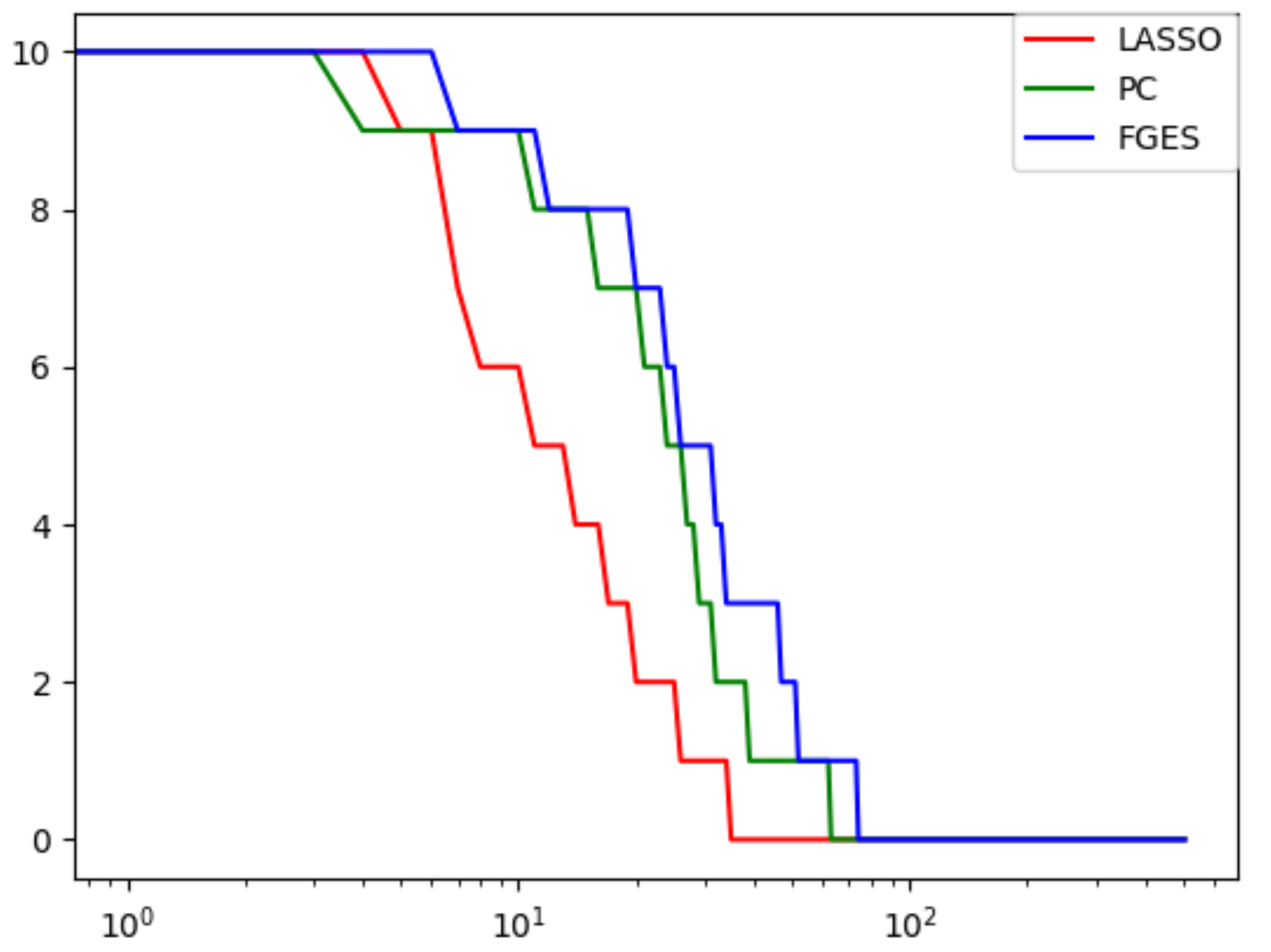}
\caption{Openness}
\end{subfigure}
\hspace{-15pt}
\label{fig:stability}
\end{figure*}

Visitors who participated in our study were asked to fill out a questionnaire, in which they were asked about the previously mentioned experiences at the resort, as well as demographical questions and the Ten-Item Personlity Inventory (TIPI). The TIPI was developed as a shortened version of a 50-item personality scale and is often used to measure Big5 personality traits. The TIPI has 10 items, with two items measuring each of the personality traits. The TIPI uses a 7-point Likert scale from 1 (``Strongly Agree") to 7 (``Strongly Disagree"). For every participant, we average the scores of the items associated with the personality to achieve five personality scores. 

A total of 3997 visitors participated. After filtering out the responses with missing values, we collected the individual characteristics of 3293 participants. Figure \ref{fig_demog} show the distribution of demographics. In the remaining data 31\% of are males and 69\% are females. The average age is 44 years old. The average number of children in the group is 1.2. Figure \ref{fig_big5} show the distribution of big5 personalities. The range of the scores are between [1,7]. Our participants experienced a total of 505 different things in the park.

The objective of the study is to measure individual differences based on visitors' experiences and choices at the theme park resort. As mentioned, studies that analyze people's digital footprints generally have an advantage in that they are more likely to reflect the choices of the individual filling out the personality survey (such as Facebook likes, publishing/commenting etc.). Most people do not visit the theme parks alone (the average group size is larger than 3 in our dataset). We have to assume that in many cases parents make decisions based on their children's preferences. That makes it difficult to link individual characteristics to the choices made by the parents. In the following section, we employ causal inference to capture effects caused by individual characteristics in order to reduce others' impacts in decision making.

%% file: causal.tex
\section*{Causal Identification}
Causal identification aims at inferring causal relationships from observational data \cite{pearl2009causality,imbens2015causal}. Many observed correlations in observational data are mediated through unobserved confounding variables. The goal of causal identification is to distinguish between such mediated correlations and truly causal relationships.

The basis of our causal identification analysis is a Bayesian network. This is a directed acyclic graph (DAG) in which nodes represent variables and arrows between the nodes represent the direction of causation between the nodes. The problem is that we only observe the nodes; these are our observations (location data, metadata, and the Big5 scores). We want to infer arrows that best explain the observed correlations. This is done in such a way that the number of causal arrows between the nodes is, in some sense, \emph{minimal} to be consistent with the observed correlations. Having identified the causal structure, we exclude all predictor variables which are not causally associated with the target variables of interest. This is the central idea of how we propose to select features.


There are two main classes of algorithms to learn causal DAGs from observational data: constraint-based and score-based ones. Constraint-based methods use independence and dependence constraints obtained from statistical tests to narrow down the candidate graphs that explain the data. 
Our first proposed method relies on using the PC algorithm \cite{spirtes1995learning} for feature selection; it belongs to the first class. The algorithm begins by learning an undirected Bayesian network that explains the variations by running conditional independence tests. The second phase consists of orienting the Bayesian network, thereby avoiding cycles and v-structures. We use a significant level of 0.05 in the independent test.



Score-based methods, on the other hand, provide a metric of confidence in the entire output model. The algorithm we use in this paper is \emph{Fast Greedy Equivalence Search} (FGES), which greedily searches over Bayesian network structures \cite{ramsey2015scaling}, and outputs the highest scoring model it finds. The score which the algorithm uses is the Bayesian Information Criterion (BIC) \cite{raftery1995bayesian}:
\begin{align*}
	BIC = 2 \cdot \ln P(data|\theta,M) - c \cdot k \cdot \ln(n).
\end{align*}
Above, $M$ denotes the Bayesian network, $\theta$ denotes its parameters, $k$ is the total number of parameters, and $n$ the number of data points. The constant $c$ is a free parameter that we can tune. This constant penalizes large numbers of parameters and thus determines the complexity of the network; we chose $c=0.1$ to obtain a comparable number of features between PC and FGES. The FGES performs a forward stepping search in which edges are added between nodes in order to increase the BIC until no single edge addition increases the score. Then, it performs a backward stepping search in which unnecessary edges are removed. 


Both the PC and FGES algorithms are used in this paper for feature selection and compared against the LASSO approach in our experimental section. The L1 term ~\footnote{We tried different values of $c=1, 0.1, 0.01$, without noticing a significant change in predictive performance.} is set as 0.01 to obtain a comparable number of features.

%% file: exp_pred.tex
\begin{table*}[t]
\centering
    \caption{Root mean square error (RMSE) and coefficient of determination ($R^2$) results for the personality prediction using visitors' choices and metadata}
    \label{tab:big5_results}
    \begin{tabular}{@{}p{1cm}p{1cm}p{1cm}p{1cm}p{1cm}p{1cm}p{1cm}p{1cm}p{1cm}p{1cm}p{1cm}@{}}
      \toprule
       Approach & Agr & & Cons & & Extr & & Neu & & Open & \\
       &   $R^2$ &RMSE &  $R^2$ &RMSE &  $R^2$ &RMSE & $R^2$ &RMSE &  $R^2$ &RMSE\\
      \midrule
       \multicolumn{2}{l}{\textbf{Visitors' choices}}\\
       \midrule
       LASSO &-0.090&0.171&-0.085&0.175&-0.087&0.242&-0.121&0.192&-0.065&0.171\\
       PC&1.143&\textbf{0.170}&\textbf{1.802}&\textbf{0.174}&1.975&0.240&\textbf{1.090}&\textbf\textbf{0.190}&\textbf{1.954}&\textbf{0.169}\\
       FGES &\textbf{1.770}&\textbf{0.170}&1.502&\textbf{0.174}&\textbf{2.368}&\textbf{0.239}&1.088&\textbf{0.190}&1.646&\textbf{0.169}\\
       \midrule
       \multicolumn{3}{l}{\textbf{Visitors' choices+Metadata}}\\
       \midrule
       LASSO &5.942&0.166&1.152&0.174&-0.338&0.242&2.109&\textbf{0.189}&1.047&0.170\\
	   PC &6.208&0.166&\textbf{3.625}&\textbf{0.172}&2.248&\textbf{0.239}&\textbf{2.959}&\textbf{0.189}&2.388&\textbf{0.169}\\
       FGES &\textbf{6.951}&\textbf{0.165}&3.525&	\textbf{0.172}&\textbf{2.482}&\textbf{0.239}&\textbf{2.995}&\textbf{0.189}&2.251&\textbf{0.169}\\
      \bottomrule
      \multicolumn{11}{l}{The five personality are \emph{Extraversion(Extr)}, \emph{Agreeableness(Agr)}, \emph{Conscientiousness(Cons)},\emph{Neuroticism(Neu)},}\\
      \multicolumn{11}{l}{and \emph{Openness(Open)}. LASSO: $\alpha = 0.01$. PC: $p=0.05$. FGES: $c = 0.1$.}
    \end{tabular}
\end{table*}
\begin{table*}
\small
\centering
    \caption{Root mean square error (RMSE) and coefficient of determination ($R^2$) results for the demographics using visitors' choices and metadata}
    \label{tab:demog_results}
    \begin{tabular}{@{}p{1cm}p{1cm}p{1cm}p{2cm}p{1cm}p{1cm}p{1cm}@{}}
      \toprule
       Approach & Age & & \# of Children & & Income & \\
       &   $R^2$ &RMSE &  $R^2$ &RMSE &  $R^2$ &RMSE\\
      \midrule
       \multicolumn{2}{l}{\textbf{Visitors' choices}}\\
       \midrule
       LASSO & 2.294&0.146&-16.53&0.153&-0.352&0.256\\
	   PC &13.675&\textbf{0.137}&0.097&0.142&\textbf{8.160}&\textbf{0.245}\\
	   FGES&\textbf{13.843}&\textbf{0.137}&\textbf{0.815}&\textbf{0.141}&8.107&\textbf{0.245}\\
       \midrule
        \multicolumn{3}{l}{\textbf{Visitors' choices+Metadata}}\\
       \midrule
       LASSO & 8.432 & 0.141 & -2.386&0.143&1.160&0.255\\
       PC & 21.684&\textbf{0.131}&8.618&\textbf{0.135}&\textbf{10.936}&\textbf{0.242}\\
       FGES & \textbf{21.975}&\textbf{0.131}&\textbf{9.539}&\textbf{0.135}&10.606&\textbf{0.242}\\
      \bottomrule
      \multicolumn{7}{l}{LASSO: $\alpha = 0.01$. PC: $p=0.05$. FGES: $c = 0.1$.}
    \end{tabular}
\end{table*} 

\begin{table*}[htb]
\centering
    \caption{Identified items that are caused by  personality found by PC algorithm. Each [.] present a facility or a meta feature which are used as features. Apart from the personal meta data, only the facility ids are used in our method. Due to confidentiality, we do not list the names of the facilities; we only show the metadata of the facilities to interpret the results. The metadata of the facilities themselves are not used in the analysis. The blue colored features indicate that the causal relationships are mutual. }
    \label{tab:Causal}
    \begin{tabular}{ p{1cm} p{15cm}} 
     \toprule
\textbf{Trait}	&	\textbf{Services/Metadata} \\
\cmidrule(lr){1-2}
Agr & [Park A: Restaurants, Quick Service],~ [Park B: Toys, Apparel, Accessories],~ [Park A: Housewares, Food],~ [Park A: Gift shop],~ [Park B: Housewares, Apparel, Accessories],~ [age],~ {\color{blue}[gender]}\\
\cmidrule(lr){1-2}
Cons & [Park A: Camera, Media, Apparel, Accessories],~ [Resort A],~ [Resort B: Spa, Pool Bars],~ [Resort B: Restaurants, Quick Service],~ [Resort B: Restaurants, Table Service],~ [Park: Restaurants, Quick Service],~ [Resort C: Apparel, Accessories],~ [Park F: Camera, Media, Apparel, Accessories],~ [Park C: Gift shop],~ [Park D: Apparel, Accessories],~ {\color{blue} [income]}\\
\cmidrule(lr){1-2}
Neu & [Park A: Restaurants, Character dining],~ [Park A: Restaurants, Table service],~ [Resort D: Restaurants, Buffet/Family Style],~ [Park A: Restaurants, Table service], [Park A: Gift shop],~ {\color{blue}[age]},~ {\color{blue}[income]}\\
\cmidrule(lr){1-2}
Extr & [Park C: Restaurant, Quick Service ],~ [Park A: Character showcase, Preschool, Kids],~ [Park C: Gift shop, Apparel, Accessories],~ [Park C: Indoor theater],~ [Park D: Gift shop, Camera, Media],~ [Resort A],~ [Resort D: Lounges],~ [Park A: Apparel, Accessories],~ [Park D: Restaurant, Character Dining, Family Style],~ [Park E: Gift shop],~ [Park A: Art Collection, Gift shop],~ [Resort E: Health beauty],~ [Park F: Gift shop, Apparel, Accessories],~ {\color{blue}[gender]},~ {\color{blue}[\# of children]} \\
\cmidrule(lr){1-2}
Open & [Park A: Thrill ride],~ [Park G: Thrill ride]~, [Park A: Thrill ride],~ [Park D: Indoor theatre],~[Park D: Restaurants, Quick Service],~[Park A: Restaurants, Quick Service],~[Park A: Restaurants, Quick Service],~ [Park A: Gift shop, Apparel, Accessories],~ [Park A: Slow ride],~ [Park D: Gift shop, Apparel, Accessories],~ {\color{blue} [age]}\\
\bottomrule
    \end{tabular}
\end{table*}

\section{Predictive Evaluation}
We implemented different prediction models to infer individual characteristics from location history and metadata. The experiments are designed to answer the following questions: a) Can the visitors' choices infer individual characteristics? b) Is there any benefit of performing feature selection according to causal identification? c) Is metadata informative to predict individual characteristics? To answer (a), we denote the visitors' choice history as a binary representation of a fixed size vector, 1 represents that the visitor visited a place. We use LASSO (Least Absolute Shrinkage and Selection Operator)\cite{tibshirani1996regression} linear regression to perform predictive tasks for continuous outcomes like age, income, the number of children and personality. LASSO is a regression method that automatically performs both feature selection and regression. All results are based on 10-fold cross validation to avoid overfitting. To answer (b),  we run PC and FGES to search for causal explanations on training data for each fold. We train predictive models by using effects of features caused by specific characteristics and predict individual characteristics on the test set. To answer (c), we add metadata in models trained in (a) and (b) to see whether the performance improves.

To identify causal relationships, we employ a Tetrad \cite{spirtes2000causation}\footnote{http://www.phil.cmu.edu/tetrad/} 
analysis to manipulate and individually study the different individual characteristics. Tetrad provides different causal search algorithms to search when there may be unobserved cofounders of measured variables and output graphical representations \cite{scheines1998tetrad}. In our experiment, we run PC and FGES search algorithms. In PC search, we choose a Fisher-z transformation to do independent tests which converts correlations into an almost normally distributed measure. The p-value is set at 0.05. In FGES setting, the penalty discount, which is $c$ in the BIC formula, is set to 0.1 in order to get a similar number of features as the PC algorithm can get from training data.

\paragraph{Stability Test} 
To compare the stability of feature selection methods, we calculated the selection probability for each feature in the Big5 personality prediction task under bootstrapping, following \cite{mandt2017sparse}. We thereby randomly removed 10 percent of the data and applied our feature selection algorithms on the remaining data. We repeated this procedure 10 times and computed the empirical feature selection probabilities. Figure \ref{fig:stability} shows the results, where the features have been ranked according to their selection probability. We see that the top 5-10 features are mostly stable across all methods, with a slightly better performance of the LASSO method on extraversion, conscientiousness and agreeableness, and with a slightly better performance of our proposed causal identification methods on openness and neuroticism.
Note that the features selected by the different methods are different. Next, we compare how well the selected features serve in our personality prediction task.


\paragraph{Predictive Performance} 
To evaluate performance, for continuous outcomes, we evaluate the results based on \textit{root mean squared error} (\textit{RMSE}) and \textit{Co-efficient of Determination}($R^2$). $RMSE$ measures the difference between values the model predicted and the observed values. $RMSE$ can be described by the following formula: 
\begin{align*}
	RMSE = \sqrt{\frac{\sum_{t=1}^{n}(y_{obs}^{t}-y_{pred}^{t})^2}{n}}
\end{align*}
where $y_{obs}^{t}$ and $y_{pred}^{t}$ are the observed and predicted scores for instance $t$, and $n$ is the sample size. $R^2$ is the ratio of the model's absolute error and the baseline mean predicted scores. It is expressed as:
\begin{align*}
	R^2 = 100 \times \left(1-\frac{\sum_{t=1}^{n}(y_{obs}^{t} - y_{pred}^{t})^2}{\sum_{t=1}^{n}(y_{obs}^{t} - \bar{y}_{obs})^2}\right)
\end{align*}
Above, $\bar{y}_{obs} = \frac{1}{n} \sum_{t=1}^{n}y_{obs}^{t}$ is the mean of the observed scores. $R^2$ contains the ratio of the variance of the prediction over the empirical variance of the scores. If this ratio is small, the prediction is accurate and $R^2$ is large. If the ratio approaches $1$, then $R^2$ approaches $0$. Negative values indicate that the prediction is not reliable. 



Usually, the metadata are informative for individual characteristics, e.g. the trip cost is correlated with income and the number of children; gender is correlated with personality. To examine the effectiveness of metadata in prediction performance, we have two datasets in the experiments: one includes only the visitors' choices, while the other one includes both the visitors' choices and metadata. Table \ref{tab:big5_results} shows the prediction performance of the Big5. The LASSO model without metadata shows visitors' choices did not capture enough predictive information, because all $R^2$  score are lower than 0. Besides, We also compare prediction performance of three feature selection methods. After employing causal identification in feature selection, the performance is higher than using LASSO. The highest $R^2$ score is obtained from FGES in predicting extraversion ($R^2 = 2.368$). The lowest RMSE is obtained from PC and FGES model in openness prediction. When the metadata are added in the models, the overall performance of all models improves, e.g. the performance of predicting agreeableness increased from $R^2=1.77$ to $R^2=6.951$ which is the best performance for the Big5.


Table \ref{tab:demog_results} shows the prediction performance of demographic information. Among the results of three outcomes (age, the number of children and income), visitors' choices are more helpful to predict age ($R^2 = 2.294$) than to predict the number of children($R^2 = -16.53$) and income($R^2 = -0.352$) in Lasso models. This means the visitors' choices are more informative to infer age. Meanwhile, after metadata are included in the model, the overall performance improves. In general, the $R^2$ results of causal identification show the models outperform the constant average baseline. The FGES model for predicting age and the number of children achieved best performance with $R^2 = 13.843$ and $R^2 = 0.815$. The PC model for predicting income achieved best performance with $R^2 = 8.160$. 

In summary, there are two findings from predictive evaluation: 1) we prove the effectiveness of the metadata in predicting individual characteristics; 2) the features selected based on causal identification can improve the prediction performance of individual characteristics.

%% file: result_analysis.tex
\section{Causal Relationship Analysis}

In addition to building models that predict individual characteristics, we are also interested in understanding the causal relationship between a visitor's characteristics and their choices. The Tetrad method specifies causal relations among the variables via a representation of a directed graph. The edge $X \rightarrow Y$ can be interpreted as X has a direct causal effect on B. We extract causal relationships between personality traits and visitors' choices with metadata from CBN, which are showed in table \ref{tab:Causal}. Within the 505 locations, agreeableness and neuroticism have direct causal effects only on 5 of the guests' choices. Agreeableness links to some specific quick food restaurants and some gift shops (we omit actual names of these locations). People who score higher agreeableness tend to visit popular parks (Park A and Park b) which are visited by most people. People who score higher on neuroticism link to some family style restaurants and dining events, which mean they tend to enjoy the comfort associated with environment. Conscientiousness has direct causal effects on 10 locations, most are facilities located in a specific hotel area of the theme park resort area, which means that conscientious people may be more likely to spend time near their hotel. Openness has direct causal effects on 10 location visits including thrill rides, quick food restaurants, and indoor theatres. Comparing with other personality traits, people who score higher openness link to most thrill rides with big drop, which mean they enjoy new experiences and seek out adventure. Extraversion has direct effects on visits of many different places (13 places, located in 6 different parks). 

%% file: discussion.tex
\section{Discussion}
We demonstrated predictive improvements by using causal identification. Personality inference is complex and can be influenced by factors such as age \cite{caspi1995temperamental}, income \cite{raadal1995prevalence}, family environment \cite{hoffman1991influence} and cultural differences \cite{hofstede2004personality}. We did not try to capture all the relevant factors in our model, instead we only used the metadata of a visitors' profile that they shared with us. However, if one were to analyze a fuller picture of visitors' choices - such as when they were where - this might show to be a more interpretable causal relationship. Meanwhile, in our model, we only use the item ids of facilities in park without exploring causal relationships between individual characteristics and the facilities' metadata. Basically any data that are a direct result of the target person's decision-making could be helpful to predict that person's characteristics. 

\section{Conclusion}
In our study, we focused on three main tasks. (1) We employed causal identification to select features which are most informative of individual characteristics (2) we built personal traits prediction models based on guests' choices and metadata (3) we employed causal relationship analysis to obtain human interpretable results. 
Our investigation has shown that models using causal identification significantly outperform the baseline models in all individual characteristics' prediction, as well as demonstrated the causal relationships between guests' choices and their personality traits.

%% file: acknowledgement.tex
\section{Acknowledgement}
We appreciate Michelle Ma for providing the illustration of an example in our paper. We are also grateful to Kun Zhang for his valuable advice and guidance about identifying causal relationship using Tetrad.